\documentclass[%
 reprint,
showpacs,preprintnumbers,
 amsmath,amssymb,
prb,
]{revtex4-1}

\usepackage{graphicx}
\usepackage{dcolumn}
\usepackage{bm}

\begin{document}


\title{\textbf{Transport and thermoelectric properties of the LaAlO$_3$/SrTiO$_3$ interface}}%

\author{A.\,Jost}
\author{V.K.\,Guduru}
\author{S.\,Wiedmann}
\author{J.C.\,Maan}
\author{U.\,Zeitler}
\email{u.zeitler@science.ru.nl}
\affiliation{High Field Magnet Laboratory, Institute for Molecules and Materials, Radboud University Nijmegen, Toernooiveld 7, 6525ED Nijmegen, The Netherlands}

\author{S.\,Wenderich}
\author{A.\,Brinkman}
\author{H.\,Hilgenkamp}
\affiliation{Faculty of Science and Technology and MESA+ Institute for Nanotechnology, University of Twente, 7500AE Enschede, The Netherlands}%

\date{\today}

\begin{abstract}
The transport and thermoelectric properties of the interface between SrTiO$_3$ and a 26-monolayer thick LaAlO$_3$-layer grown at high oxygen-pressure have been investigated at temperatures from 4.2\,K to 100\,K and in magnetic fields up to 18\,T. For $T>$\;4.2\,K, two different electron-like charge carriers originating from two electron channels which contribute to transport are observed. We probe the contributions of a degenerate and a non-degenerate band to the thermoelectric power and develop a consistent model to describe the temperature dependence of the thermoelectric tensor. Anomalies in the data point to an additional magnetic field dependent scattering.
\end{abstract}

\pacs{74.40.-c, 72.15.Jf}
\maketitle
\section{Introduction}
Since the discovery of conduction at the interface between the two band-insulating perovskite oxides SrTiO$_3$ and LaAlO$_3$\cite{Ohtomo2004} a plethora of new effects have been found, ranging from superconductivity\cite{Reyren2007} to magnetism\cite{Brinkman2007,BenShalom2009, Ariando2011, Dikin2011, Li2011, Bert2011, Lee2013, Huijben2006} and tunable switching of high mobility interface conductivity,\cite{Thiel2006, Caviglia2008, Guduru2013} depending on the ground state the sample reaches. The nature of the ground state present in the systems depends closely on the growth parameters of the LaAlO$_3$-layer,\cite{Brinkman2007} the LaAlO$_3$ layer thickness\cite{Kozuka2010, Wong2010, Bell2009} and configuration of the heterostructures with different capping layers on top of LaAlO$_3$.\cite{Huijben2006, McCollam2014, Huijben2009, Huijben2013}\\
A number of possible mechanisms are proposed to be responsible for the conduction at the interface,\cite{Nakagawa2006, Yoshimatsu2008, Sing2009, Siemons2007, Herranz2007, Willmott2007} which can lead to multiple charge carrier conduction if several are active. Indeed multiple carrier conduction has been found in SrTiO$_3$/LaAlO$_3$-interfaces over a wide range of growth condictions.\cite{McCollam2014, Guduru2013a,Popovic2008a, Lerer2011,BenShalom2009, Rakhmilevitch2013, Seo2009}\\
Until present, most investigations on the electronic properties of LaAlO$_3$/SrTiO$_3$ interfaces have been done using transport experiments,\cite{Hwang2012, Mannhart2010, Ohtomo2004, Reyren2007, Brinkman2007, BenShalom2009, Ariando2011, Dikin2011, Li2011, Bert2011, Lee2013, Huijben2006, Thiel2006, Caviglia2008,Kozuka2010, Wong2010, Huijben2013, McCollam2014, Huijben2009, Guduru2013, Guduru2013a, BenShalom2009, Rakhmilevitch2013, Seo2009, Bell2009} while measurements of thermoelectric power are still sparse in SrTiO3/LaAlO3-interfaces.\cite{Pallecchi2010, Lerer2011, Filippetti2012}  Whereas transport experiments are generally dominated by the charge carrier mobility, contributions of lower mobility can be accessed in thermoelectric power measurements. Additionally, the thermoelectric power is known to be sensitive to magnetic scattering, thus an ideal tool to investigate multiple charge carrier contributions in samples with magnetic signatures.\\
In the present work we report on our investigation of the interface electronic structure of one specific type of LAO/STO heterostructure, with a 10\,nm (26 unit cells) LAO film, which is known to exhibit magnetic signatures.\cite{Brinkman2007} Magnetotransport and thermoelectric-power measurements have been performed in a large temperature and magnetic field range. We apply a two carrier model to the magnetotransport data and find two different charge carriers with different densities and mobilities. By combining transport and thermopower data, we extend this model to the thermoelectric tensor $\epsilon$ at zero field and develop a preliminary description for its behaviour in magnetic field.\\
The paper is organized as follows: in section II we describe the details of sample growth and the experimental setup. In section III we show our transport results, which we describe in terms of a classical two-carrier model. In section IV we present our results of the thermopower-measurements and extend our model to thermopower. In section V we discuss our results and draw conclusions in section VI.

\section{Experimental details}
The sample is grown by pulsed laser deposition and has a 10\,nm thick (26 unit cells) LAO film on a TiO$_2$-terminated single crystal STO [001] substrate (treatment described in Ref. \onlinecite{Koster1998}). The LAO film was deposited at a substrate temperature of 850$\,^{\circ}\mathrm{C}$ and an oxygen pressure of $2\cdot10^{-3}$\,mbar, in order to minimize oxygen vacancies, using a single-crystal LaAlO$_3$ target. The growth of the LAO film was monitored using in-situ reflection high-energy electron diffraction. After the growth, the samples were cooled to room temperature at the deposition pressure.\\
Electrical contacts to the sample were made using an ultrasonic wire-bonder to punch through the LAO-layers and attaching manganine-wires to the holes with silver-paint.\\
The thermoelectric power was measured in a home-build apparatus in a standard one heater, two thermometer-geometry similar to the one used in Ref. \onlinecite{Fletcher1994}. The resistivity was measured on the same sample and in the same apparatus using a conventional low-noise lock-in technique. Both quantities were measured in positive and negative field directions and the data shown is symmetrized (antisymmetrized) to obtain the diagonal (off-diagonal) components of the resistivity- and thermoelectric power tensors. We use the historical sign convention for the Nernst-Ettingshausen effect throughout the paper (positive Nernst signal along $y$-direction, when the field is in $z$-direction and gradient in $x$-direction).

\section{Magnetotransport}
\label{sec:magnetotransport}
\begin{figure}[htb]
 \centering
\includegraphics[width=\linewidth]{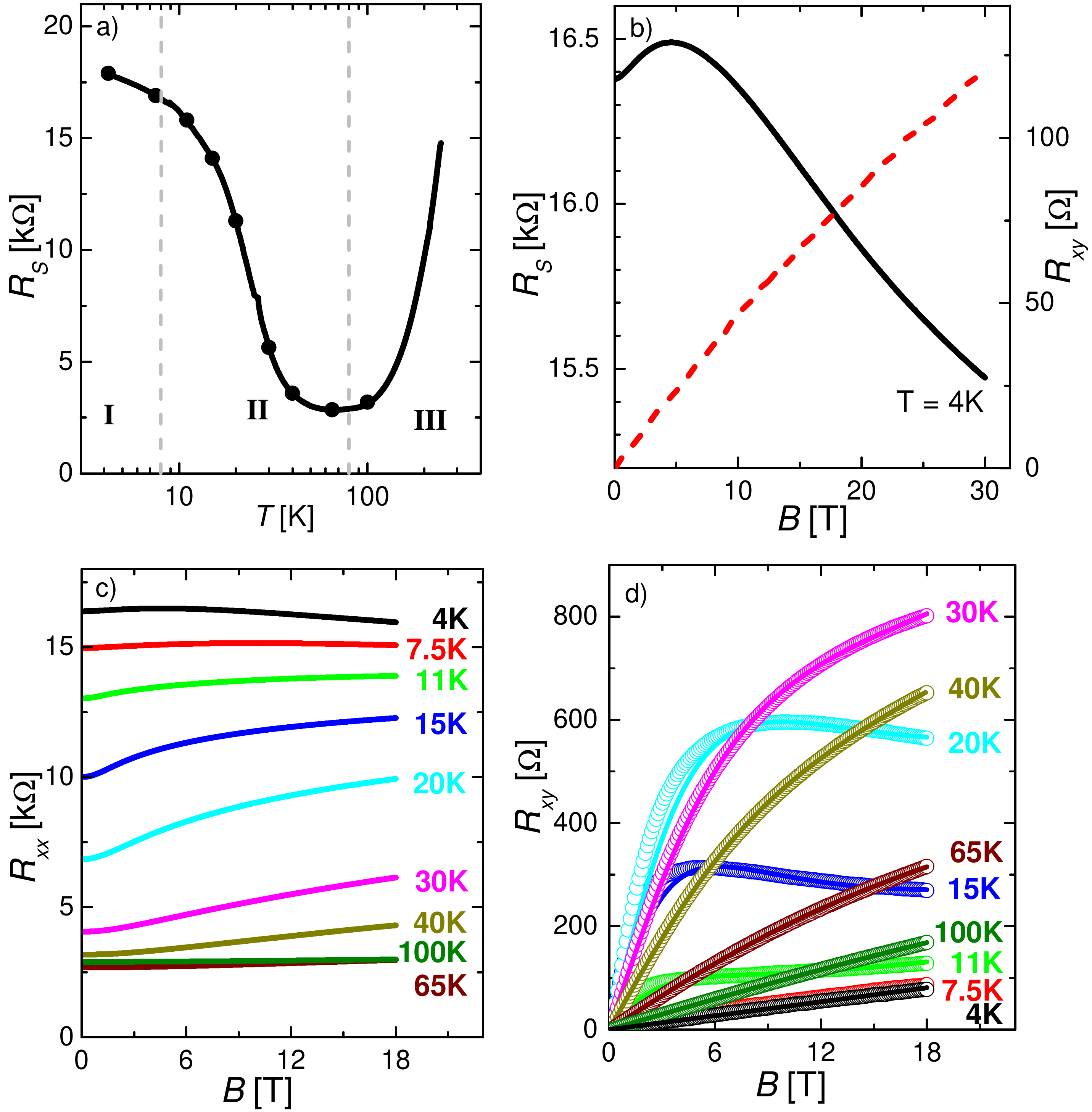}%
 \caption{(color online) a) Temperature dependence of sheet resistance (line). Sheet resistance from two-charge carrier model (dots).\\
 b) Sheet resistance (black line) and Hall-resistance (dashed red line) at 4\,K showing the negative magnetoresistance and the linear Hall-effect.\\
 c) Measured sheet resistance $R_{xx}$ for temperature between 4\,K and 100\,K.\\
 d) Measured Hall resistivity for temperature between 4\,K and 100\,K (empty dots). Fits with the two-charge-carrier model (lines).}%
\label{fig:figure1}
\end{figure}

The temperature dependence of the sheet resistance $R_S$, the magnetoresistance $R_{xx}$ and the Hall-data $R_{xy}$ are shown in figure \ref{fig:figure1}a, c and d, respectively. The analysis of the transport data follows closely our previous work \cite{Guduru2013a}. From the transport data (figure \ref{fig:figure1}), we can distinguish three different regions (from low to high temperature): region I (up to $\sim$\,8\,K) with logarithmically decreasing sheet resistance and linear Hall effect, region II ($\sim$\,8\,K to $\sim$\,50\,K) with strongly decreasing sheet resistance and strongly non-linear Hall-resistance and region III (above $\sim$\,50\,K) showing an increase in the sheet resistance and a linear Hall-effect with respect to the magnetic field.
The transport data can be analyzed using a two-charge-carrier model for two independent, electron-like channels. For T\;$>$\;11\,K the magnetic field dependent Hall-resistance $R_{xy}$ was fitted with a restriction to the zero-field resistance $R_{S0}$ using the equations\cite{AshcroftMermin1976}:
\begin{equation}
\label{eq:2chargecarrier}
		\begin{aligned}
		R_{xy} &= 	
\frac{B}{e}\frac{(n_1\mu_1^2+n_2\mu_2^2)+(\mu_1\mu_2B)^2(n_1+n_2)}{
(n_1\mu_1+n_2\mu_2)^2+(\mu_1\mu_2B)^2(n_1+n_2)^2}\\
		R_{S0}&=\frac{1}{e(n_1\mu_1+n_2\mu_2)}
		\end{aligned}
\end{equation}
with the $e$ the electron charge, $B$ the magnetic field and the independent fitting parameters $n_{1,2}$ and $\mu_{1,2}$ as charge carrier density and mobility for the two channels, respectively. For the lowest and highest 
temperature, where the Hall-resistance is completely linear, the density for one type of charge carrier was obtained from a linear fit to $R_{xy}=B/(en)$ and the mobility from $R_{S0}=1/(en\mu)$.
\begin{figure}[htb]
 \centering
\includegraphics[width=0.8\linewidth]{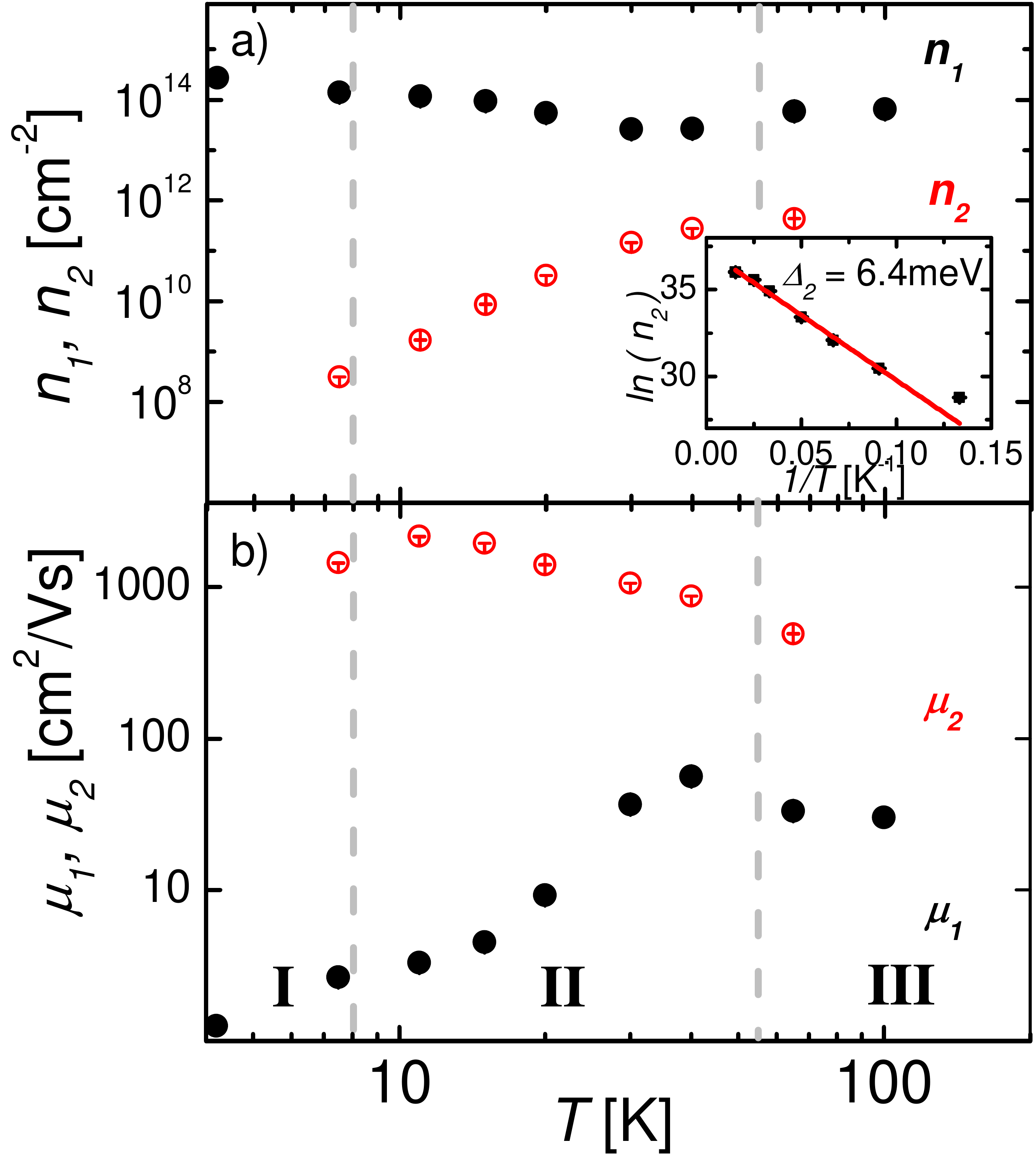}%
 \caption{(color online) a) Charge carrier densities of the two electron-like bands $n_1$, $n_2$ from the 2-charge-carrier model (inset: Arrhenuis-plot for the n$_2$).\\
 b) Mobilities of the two electron-like bands $\mu_1$, $\mu_2$ from the 
2-charge-carrier model.
}%
\label{fig:figure2}
\end{figure}
The obtained fitting parameters are shown in figure \ref{fig:figure2}.\footnote{The data-point at 7.5\,K violates the Mott-Ioffe-Regel criterion ($k_fl<1$). We still included it in the figure for completeness.} The first type of charge carriers ($n_1$, $\mu_1$) has a relatively high and temperature independent charge carrier density of about $n_1 \sim 10^{14}$\,cm$^{-2}$. The mobility of these charge carriers is low with $\mu_1 \sim 40$\,cm$^{2}/$Vs and increases at higher temperatures to $\mu_1 \sim 1$\,cm$^{2}/$Vs at 4\,K (therefore referred to as low mobility charge carriers). This decrease in mobility to lower temperatures was attributed to magnetic, Kondo-like scattering due to the negative magneto-resistance at low temperatures and is described in more detail in our previous work\cite{Brinkman2007, Guduru2013a}. The slight decrease in mobility at higher temperatures is probably due to electron-phonon scattering.\\
The second type of charge carriers has a much lower charge carrier density which is increasing with a thermally activated behaviour $n_2 \propto e^{-\Delta_2/k_BT}$ from $n_2 \sim 2\cdot10^{9}$\,cm$^{-2}$ at 11\,K to $n_2 \sim 4\cdot10^{11}$\,cm$^{-2}$ at 65\,K, i.e. we deal with charge carrier, that have to be treated as a non-degenerate electron gas. To derive the energy gap between the non-degenerate, high mobility band at higher energy and the low lying, low mobility band, we can use an Arrhenius-plot (inset of figure \ref{fig:figure2}a)), which gives a gap of $\Delta_2 = (6.4\pm0.4)$\,meV, comparable to the results on other samples\cite{Guduru2013a, Huijben2006, Huijben2013}. The mobility is high at low temperatures ($\mu_2 \sim 2000$\,cm$^{2}$/Vs) and decreases at higher temperatures by one order of magnitude (high mobility charge carriers).

\section{Thermoelectric Power}
\begin{figure}[htb]
 \centering
\includegraphics[width=0.8\linewidth]{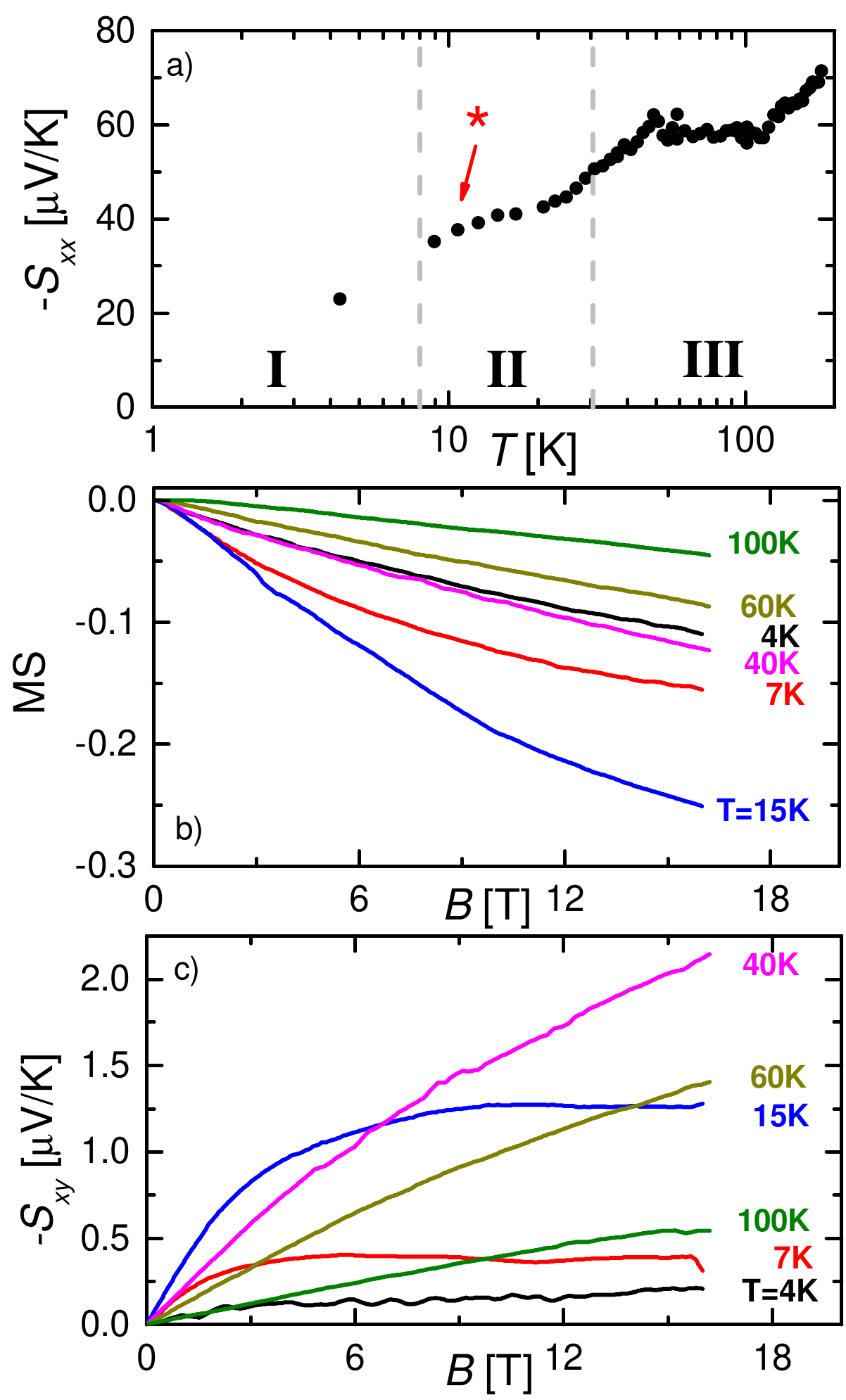}%
 \caption{(color online) a) Zero-field thermoelectric power $S_{xx}$ in a semi-logarithmic plot for clarity. A small deviation from the observerd $T^{0.4}$ dependence around 15\,K is marked by an asterisk *.\\
 b) Magnetothermopower \textit{MS} (see text) for selected temperatures.\\
 c) Nernst effect $S_{xy}$ for selected temperatures.
}%
\label{fig:figure3}
\end{figure}

\begin{figure}[ht]
 \centering
\includegraphics[width=0.94\linewidth]{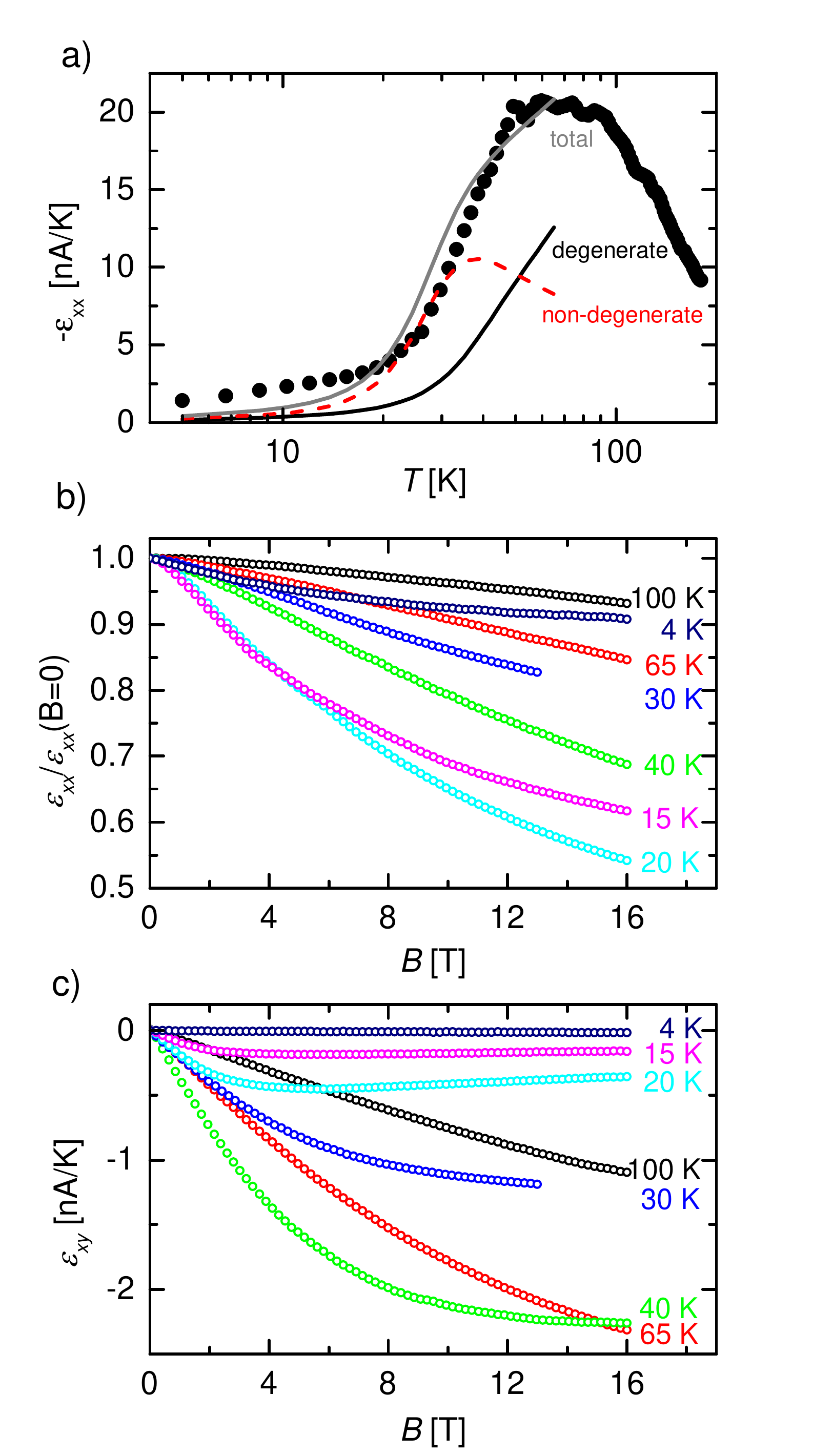}%
 \caption{(color online) a) Temperature dependence of $\epsilon_{xx}$ (black dots). For illustration typical curves for degenerate and non-degenerate electrons are shown (see text).\\
 b) Magnetic field dependence of $\epsilon_{xx}$ for selected temperatures between 4\,K and 100\,K. \\
 c) Magnetic field dependence of $\epsilon_{xy}$ for selected temperatures between 4\,K and 100\,K.\\
}%
\label{fig:figure4}
\end{figure}

The temperature dependence of the zero-field thermoelectric power $-S_{xx}$ is shown in figure \ref{fig:figure3}a. First, the thermoelectric power is negative, confirming that our carriers are electrons. Second, we point out the absence of a clear phonon-drag peak.\cite{Fletcher1994} Only a faint deviation from the observed dependence is visible around 15\,K (marked by an asterisk *). Thus we assume that we are in the regime of diffusion-dominated thermopower. The three regions identified in the transport measurements are indicated by the dashed lines. Region I cannot be clearly identified in thermopower due to the lack of data-points. In region II, the thermopower is increasing monotonously, proportional to approximately between $T^{0.3}$ and $T^{0.4}$. We note that the temperature dependence is also in agreement with thermoelectric power by variable range hopping which gives a $T^{1/3}$ dependence for a 2D electron gas.\cite{Burns1985}\\
If multiple charge carriers are present, the individual contributions of the charge carrier have to be weighted by their conductivities and we can write for $S_{xx}$ at zero magnetic field\footnote{for the thermoelectric tensor we can write as explained below: $\epsilon_{xx} = \epsilon_{xx}^{(1)} + \epsilon_{xx}^{(2)}$. With $\epsilon_{xx}^{(i)} = \sigma_{xx}^{(i)}S_{xx}^{(i)}$ we get immediately $S_{xx}(\sigma_{xx}^{(1)} + \sigma_{xx}^{(2)}) = \sigma_{xx}^{(1)}S_{xx}^{(1)} + \sigma_{xx}^{(2)}S_{xx}^{(2)}$ and equation (\ref{eq:2bandTEP}) follows.}:                                                                                                                                                                                                                                                                                                                                                                                                                                                                                                                                                                                                                                                                                                                                                                                                                                                                                                                                                                                                                                                                                                                                                                                                                                                                                                                                                                                                                                                                                                                                                                                                                                                                                                                                                                                                                                                                                                                                                                                                                                                                                                                                                                                                        
\begin{equation}
\label{eq:2bandTEP}
 S_{xx} = \frac{\sigma_{xx}^{(1)} S_{xx}^{(1)} + \sigma_{xx}^{(2)}S_{xx}^{(2)}}{\sigma_{xx}^{(1)}+\sigma_{xx}^{(2)}}.
\end{equation}
The combination of both contributions explains the observed temperature dependence and will be described later in this paper using the thermoelectric tensor (see figure \ref{fig:figure4}).\\
In region III, the thermopower is constant up to about 120\,K and increases monotonously at higher temperatures, again proportional to approximately $T^{0.4}$.
\\
When a magnetic field is applied, the magnitude of the thermoelectric power decreases. We can define the magneto-thermopower MS as
\begin{equation}
MS(B)=\frac{S_{xx}(B)-S_{xx}(B=0)}{S_{xx}(B=0)},
\label{magneto-TEP}
\end{equation}
shown in figure \ref{fig:figure3}b. In region I (4\,K), the magneto-thermopower is weak and linear with a decrease by about 11\% at 16\,T. In region II, the MS becomes stronger and non-linear with a maximal suppression of 25\% at 15\,K. Above 15\,K the magneto-thermopower becomes weaker again but remains non-linear. In region III, the magneto-thermopower is reduced even more and becomes linear again.\\
The Nernst effect $S_{xy}$ (figure \ref{fig:figure3}c) shows similar behaviour. In region I it is nearly linear and small, with $\nu = S_{xy}/B \approx -12$\,nV/(TK). We note that this is still enhanced compared to the classical Fermi-liquid picture\cite{Behnia2009}, which is approximately:
\begin{equation}
 \nu=-\frac{\pi^2}{3}\frac{k_B}{e}\frac{k_BT}{\epsilon_F}\mu \approx -[0.1...1] \frac{\mbox{nV}}{\mbox{TK}}
\end{equation}
depending on the actual Fermi-energy.
At higher temperature (region II), $S_{xy}$ becomes non-linear and increases by an order of magnitude at 40\,K and 16\,T in comparison to the value at 4\,K. When entering region III, the Nernst signal becomes linear again and decreases towards higher temperatures.\\
The overall behaviour of the thermoelectric signals in magnetic field resembles the transport signals, i.e. linear and small in regions I and III and non-linear and large in region II. It is therefore beneficial to derive an appropriate two-charge-carrier model for the thermopower, which can be done following the work of S. Cao \textit{et al.}\cite{Cao1996a}\\
The thermopower tensor is defined by $S=E/(\nabla T)$ under the condition of $J_q =0$, with $E$ the electric field, $J_q$ the charge current density and $\nabla T$ the temperature gradient. We can use an extended Ohm's law
\begin{equation}
J_q = \sigma E - \epsilon \nabla T
\end{equation}
with $\sigma$ the conductivity tensor, $\epsilon$ the thermoelectric tensor and $\rho=\sigma^{-1}$ the resistivity, to find an expression for the thermopower tensor:
\begin{equation}
\label{eq:thermopower}
S = E/(\nabla T)= \sigma^{-1}\epsilon = \rho\epsilon
\end{equation}
By rewriting equation (\ref{eq:thermopower}), we get $\epsilon = \sigma S$ for the thermoelectric tensor. Using $\sigma$ and $S$ from our measured data we obtain $\epsilon_{xx}$ and $\epsilon_{xy}$, shown in figure \ref{fig:figure4}.\\
\noindent The temperature dependence of the thermoelectric tensor $\epsilon_{xx}$ at zero magnetic field (figure \ref{fig:figure4}a) is characterized by a negative quantitity (typical for electrons), which magnitude increases linearly with temperature for T\;$<$\;25\,K. At 25\,K, the thermoelectric tensor starts to increase rapidly up to a plateau, which extends between 50\,K and 90\,K. This increase originates from two simultaneous effects: the increase in mobility of the low mobility (degenerate) charge carriers and the thermally activated population of the high mobility (non-degenerate) band. Above 90\,K, the thermoelectric tensor starts to decrease, probably due to an increase in phonon scattering.\\
To describe the zero-field data, the contributions of the degenerate and the non-degenerate band have to be added up: $\epsilon = \epsilon_d + \epsilon_{nd}$. The degenerate, low mobility band can be described by the the well-known Mott-formula:\cite{Mott1971}
\begin{equation}
\epsilon_d=-\sigma_d\frac{\pi^2k_B}{3e}\frac{k_BT}{\epsilon_F}(1+p)
 \label{eq:degenerate}
\end{equation}
with the Fermi-energy $\epsilon_F$ and the scatter parameter $p=(\partial ln\,\tau/\partial ln\,\varepsilon)|_{\varepsilon_F}$. The temperature dependence of the thermopower of a non-degenerate twodimensional electron-gas (as are the thermally excited high mobility charge carriers) is given by:\cite{Ziman1960}
\begin{equation}
 \epsilon_{nd} = \sigma_{nd}\frac{k}{e}\left[(p + 2) + \frac{\epsilon_e}{kT} \right]
 \label{eq:non-degenerate}
\end{equation}
with $\epsilon_e$ the Fermi-energy measured from the lower edge of the conduction band and $p=(\partial ln\,\tau/\partial ln\,\varepsilon)$. Thus, the thermopower of a non-degenerate electron gas is proportional to $1/T$.\\
To illustrate the behaviour of the temperature dependence of these bands, we plotted typical curves for the thermoelectric tensor of degenerate and non-degenerate electron systems $\epsilon_d$ and $\epsilon_{nd}$ in figure \ref{fig:figure4}a using the conductivities of the two bands as determined by the transport measurements.\\
As parameters for the curves we used $p=-0.5$, which is the theoretical value for hard sphere scattering,\cite{Gallagher1992} $\epsilon_e = 6.4$\,meV for the non-degenerate gas, as extracted from the thermal activation analysis,  and $\epsilon_F=20$\,meV for the degenerate electron gas, in order to archieve the right magnitude compared to the data measured. We note that $\epsilon_F=20$\,meV is in the right order of magnitude for a 2D electron gas with the density measured. Fitting the temperature dependence is not possible because the scatter parameters and energies in equations (\ref{eq:degenerate}) and (\ref{eq:non-degenerate}), respectively, are not independent from each other. The actual values of the scatter parameter $p^{(i)}$ and the energies $\epsilon_e$ and $\epsilon_F$ can be determined by magnetic field dependent measurements, using an appropriate model for the magnetic field dependence of the two bands.\\ 
The magnetic field dependence of the diagonal ($\epsilon_{xx}/\epsilon_{xx}(B=0)$) and off-diagonal ($\epsilon_{xy}$) components of the thermoelectric tensor are shown in figure \ref{fig:figure4}b and c, respectively. The magnitude of the diagonal component of the thermoelectric tensor $\epsilon_{xx}/\epsilon_{xx}(B=0)$ decreases with field at all measured temperatures. However the details of their magnetic field dependence change drastically. At low temperatures, $\epsilon_{xx}/\epsilon_{xx}(B=0)$ decreases steeply at low fields, turning into saturation towards higher fields. The total decrease is thereby the largest at 20\,K. Starting from 30\,K, $\epsilon_{xx}/\epsilon_{xx}(B=0)$ is flat at low fields, getting steaper towards high fields and saturate again at highest fields. This saturation vanishes at 100\,K within the measured field range.\\
The off-diagonal component $\epsilon_{xy}$ increases linearly at low fields, changing with a kink to a lower, still linear slope at higher fields. The field where the slope changes, increases linearly with temperature from about 2\,T at 4\,K to 4.2\,T at 20\,K. Due to the sharpeness of the kink and his $B$-$T$-dependence, we attribute this kink to be a remnant of magnetic scattering. At 30\,K and 40\,K, a similar change in the slope exists, but at higher fields and smeared out over several Teslas. Therefore, we attribute this behaviour to the existence of two different types of charge carriers. At 65\,K and 100\,K, the transition either vanishes or is so much broadened and shifted to higher fields, that it is not visible anymore in our measurement.\\
To model the field dependence of the thermoelectric tensor of the first type charge carriers, we can use an expression for the diffusion thermoelectric tensor in the classical, degenerate limit given by:\cite{Fletcher1994}
\begin{equation}
\label{eq:thermotensor}
	\begin{aligned}
	\epsilon_{xx} &= 
-\sigma_{xx}\frac{L_0eT}{\varepsilon_F}\left[1+p\frac{1-\mu^2B^2}{1+\mu^2B^2}
\right]\\
	\epsilon_{xy} &= 
-\sigma_{xy}\frac{L_0eT}{\varepsilon_F}\left[1+\frac{2p}{1+\mu^2B^2}\right]
	\end{aligned}
\end{equation}
with $L_0=\pi^2k_B^2 /3e^2$ the Lorenz-number, $k_B$ the Boltzmann-constant, $\varepsilon_F$ the Fermi energy, $p=(\partial ln\,\tau/\partial ln\,\varepsilon)|_{\varepsilon_F}$ and $\tau$ the transport lifetime. The conductivities $\sigma_{xx}$ and $\sigma_{xy}$ are calculated with 
\begin{equation}
\label{eq:sigma}
	\begin{aligned}
	\sigma_{xx} &= \frac{ne\mu}{1+\mu^2B^2}\\
	\sigma_{xy} &= \frac{ne\mu^2B}{1+\mu^2B^2}
	\end{aligned}
\end{equation}
using the charge-carrier densities $n$ and the mobilties $\mu$ obtained from the transport data (figure \ref{fig:figure2}).\\
The second type of charge carriers ($n_1$, $\mu_1$) is non-degenerate. Therefore it cannot be modeled with the same equations. To the best of our knowledge, a theoretical model for the magnetic field dependent thermopower of a non-degenerate electron gas is still missing. A development of such a model is beyond the scope of this work and remains as a future challenge. 

\section{Discussion}
The main part of our data, i.e. for temperatures above 30\,K, seems to be well described by a two-charge carrier model, also for thermoelectric power. However, below 30\,K some peculiar features are observed: first, the magnitude of the Nernst-signal is enhanced compared to the Fermi-liquid picture. Second, the kink observed in the off-diagonal component of the thermoelectric tensor $\epsilon_{xy}$, which has a linear $B$-$T$-dependence.\\
A similar anomaly is observed in transport measurements at the same temperature, namely the observation of a negative magneto-resistance shown in figure \ref{fig:figure1}b and described elsewhere.\cite{Guduru2013a, Brinkman2007} There, the negative magneto-resistance is attributed to magnetic, Kondo-like scattering. Indeed, the thermoelectric power in Kondo-lattices shows similar behaviour as in the LaAlO$_3$/SrTiO$_3$ interface.\\
In three-dimensional Kondo-lattices with one type of charge carrier, the Seebeck coefficient $S_{xx}$ is decreasing strongly with magnetic field and saturating at high fields, and the Nernst coefficient $\nu=S_{xy}/B$ is large at small fields and decreasing to higher fields.\cite{Bel2004, Falkowski2011} In other words, the Nernst effect is increasing strongly at low magnet fields and saturates at high fields. In our measurements, we observe the same behaviour for low temperatures; a strong decrease in the Seebeck effect with saturation to high fields and a strong increase in Nernst at low field saturating at high fields (see figure \ref{fig:figure3}). Since we observe the same signatures as in Kondo-lattice materials, we suggest that a similar mechanism could play a role in our sample and we attribute the strong magnetic field dependence of the thermoelectric tensor to an additional magnetic scattering acting on the low mobility charge carriers ($n_1, \mu_1$) at low temperature.\\
A possible route to magnetic scattering can be explained by the polar-catastrophe scenario,\cite{Nakagawa2006} where charge is transfered to the interface due to the polarity of the LaAlO$_3$-layers. This additional charge can change the electric state of the non-magnetic Ti$^{4+}$-ions in SrTiO$_3$ to magnetic Ti$^{3+}$-ions.\cite{Pentcheva2007, Lee2013} These magnetic Ti$^{3+}$-ions then can act as scattering-partners for the electrons at the interface and the low mobility charge carrier are located close to the interface at the Ti$^{3+}$-ions. Over the location of the high mobility charge carriers we cannot give any conclusion.\\
We lack an appropriate model to determine the density of magnetic moments at the interface from our data. However, we can place an upper limit by assuming that they are created by the electrons arising from the polar-catastrophe. This would give an upper limit of $1/2$ a moment per unit cell or $3.4 \times 10^{-14}$\,cm$^{-2}$.

\section{Conclusion}
We have measured a complete set of transport and thermoelectric power data in a temperature range from 4\,K to 100\,K in fields up to 18\,T. We find two different electron-like charge carrier with different densities and mobilities: a degenerate band, with a low mobility and a high carrier density and a non-degenerate band with a higher mobility, which vanishes at low temperatures. The temperature dependence of the thermoelectric tensor can be described by this two-band picture, but for the magnetic field dependence an appropriate model for the non-degenerate band is still missing. We identify anomalies in the thermopower data, which cannot be readily explained by the two-band picture. We attribute them to an additional strongly magnetic field dependent scattering mechanism of the low mobility charge carriers located close to the Ti$^{3+}$-atoms at the interface. 

\section{Acknowledgments}
This work has been performed at the HFML-RU/FOM member of the European Magnetic Field Laboratory (EMFL) and is part of the InterPhase research program of the Foundation for Fundamental Research on Matter (FOM, financially supported by the Netherlands Organization for Scientific Research (NWO)).

\bibliography{used_in_papers-LAO-STO}
\end{document}